\documentclass[aps,pre,reprint,twocolumn,groupedaddress,showpacs]{revtex4}
\usepackage{color}
\usepackage{graphicx}
\usepackage{bm}

\usepackage{amsbsy}
\usepackage{amssymb}

\def\dd{\mbox{d}}

\begin{document}

\title{Coarsening and accelerated equilibration in
  mass-conserving heterogeneous nucleation}

%
%

\author{Tom Chou$^{1}$ and Maria R. D'Orsogna$^{2}$}
\affiliation{$^{1}$Depts. of Biomathematics and Mathematics, UCLA, Los Angeles, CA 90095-1766}
\affiliation{$^{2}$Department of Mathematics, CSUN, Los Angeles, CA 91330-8313}

\date{\today}




\begin{abstract}

We propose a model of mass-conserving heterogeneous nucleation to
describe the dynamics of ligand-receptor binding in closed cellular
compartments. When the ligand dissociation rate is small, competition
among receptors for free ligands gives rise to two very different
long-time ligand-receptor cluster size distributions.
Cluster sizes first plateau to a long-lived,
initial-condition dependent, ``metastable'' distribution, and coarsen
only much later to a qualitatively different equilibrium one.
Surprisingly, we also find parameters for which a very special subset
of clusters have equal metastable and equilibrium sizes, appearing to
equilibrate much faster than the rest.  Our results provide a
quantitative framework for ligand binding kinetics and suggest a
mechanism by which different clusters can approach their equilibrium
sizes in unexpected ways.

\end{abstract}


\maketitle

\section{Introduction}

The binding of multiple particles to specific nucleation sites is a
key process in many physical and chemical settings.  The formation of
droplets, condensates on aerosols \cite{KROON2011,LAZARIDIS2000}, and crystals
\cite{WINKLER2008} is often triggered by the presence of impurities or
boundaries, in a process known as heterogeneous nucleation
\cite{DJIKAEV2005,GMBH}.  Heterogeneous nucleation also occurs in cell
biology during the assembly of sickle hemoglobin \cite{FERRONE},
$\beta$-amyloid fibers \cite{LOMAKIN}, Arp2/3 complex-mediated actin
nucleation \cite{DAYEL}, and probably during clathrin-coat assembly
\cite{CLATHRIN}.  Within biochemical applications, ligand-receptor
binding can also be viewed as a particular paradigm of heterogeneous
nucleation, where multiple ligands bind to a single receptor akin to
an impurity seed in solid-state nucleation.

Viewed through this lens, nucleation is ubiquitous in cell
biology. Indeed, receptor loading levels control a variety of
biochemical reactions, from viral entry to cell signalling.  The
chemical stoichiometries involved in ligand-binding events however,
may limit the maximum number of ligands a receptor can hold to about a
dozen.  For example, hemoglobin can bind at most four oxygen molecules
\cite{WALSH}, virus-cell fusion occurs after a small number of cell
surface receptors bind to a viral protein \cite{OURJCP}, and
cell-signaling is initiated after a certain number of phosphates bind
to specific enzymes \cite{PHOS}.  This is in contrast to most physical
and chemical systems where aggregation of an unlimited number of
particles can lead to the emergence of macroscopic structures.

\begin{figure}[t]
\begin{center}
\includegraphics[width=3.3in]{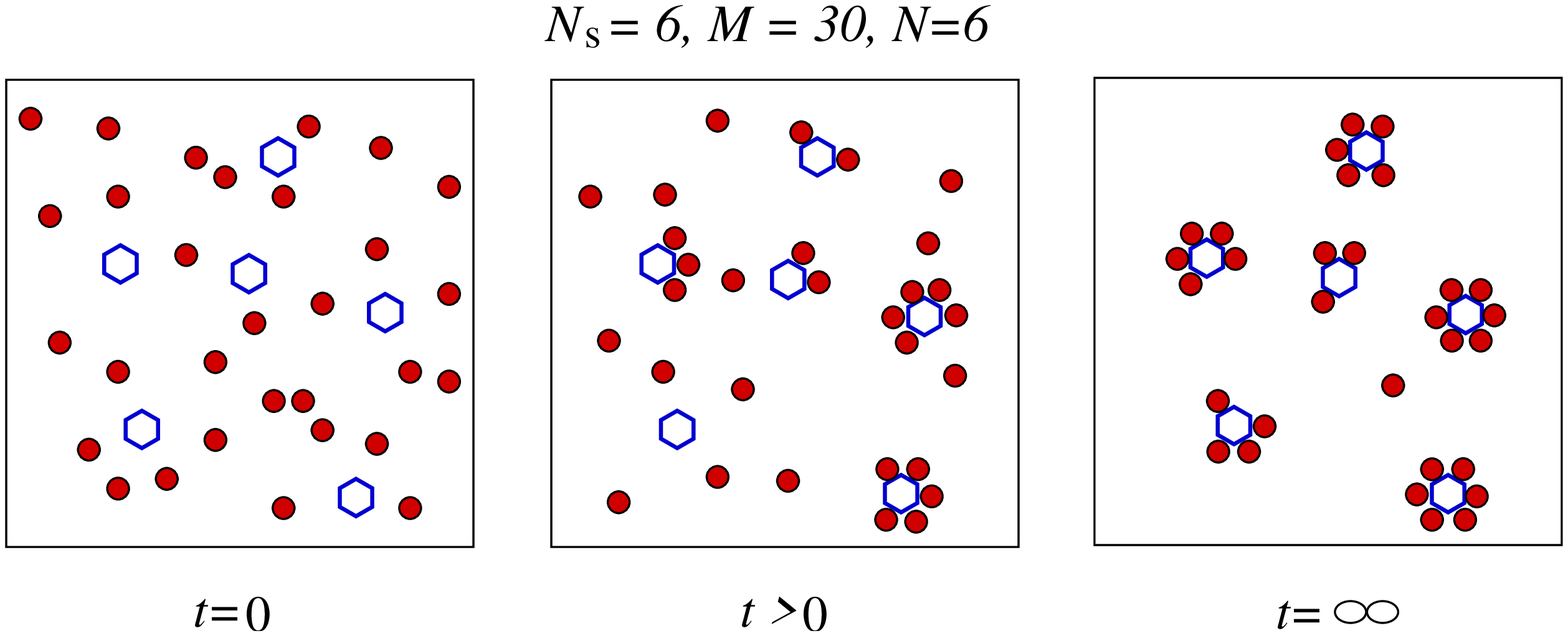}
\caption{A heterogeneous nucleation process in which ligand monomers bind only to
seeds. Here, $N_{\rm s} = 6$ seeds (open hexagons) are available to
bind $M=30$ initial monomers (filled dots).}
\label{FIG1}
\end{center}
\end{figure}

An even more critical feature of nucleation in cellular settings is
the small system volumes involved and, as a consequence, the presence
of a finite number of monomeric ligands driving the nucleation
process.  In the small volumes encountered in cells, ligand production
and degradation are often slower processes than attachment and
detachment to receptors, allowing certain ligands to be depleted
\cite{ZIA2}.  Because there is no source to replenish the free ligand
concentration, receptors in confined, isolated systems compete amongst
themselves for the finite pool of free monomeric ligands, as depicted
in Fig.~\ref{FIG1}.  For simplicity, throughout the remainder of this
paper, the terms ``monomers'' and ``ligands'' will be used
interchangeably, as will ``seeds'' and ``receptors.''

The dynamics of mass-conserving {\it homogeneous} nucleation has been
well-studied in the context of Becker-D\"{o}ring equations
\cite{WATTIS,PENROSE,REDNER}. In this work we study its
\textit{heterogeneous} counterpart, relevant for ligand-receptor
kinetics in biology.  While heterogeneous nucleation has been well
studied, most theoretical treatments focus on computing equilibrium
partition functions for nucleation with specific forms for the free
energy of monomer association
\cite{LAZARIDIS1993,FLETCHER1958,HALE1974}.  Many other approaches
focus on either the molecular details and geometry of an individual
cluster particle \cite{NOWAKOWSKI1992}, or on the asymptotic dynamics
of even more coarse-grained continuum size distributions
\cite{NEU}. In many applications, a constant source of monomers is
also imposed \cite{BHATT2003}. Here, we will instead consider the
dynamics of a system with a total fixed number $M$ of monomeric
ligands (bound and unbound) and a fixed number $N_{\rm s}$ of receptor
seeds. Each receptor can bind at the most $N$ monomeric ligands
according to the spatially-uniform mass-action equations we describe
in the next section.

Two qualitatively different cases are analyzed. In Section III, we
first consider irreversible binding, where the detachment rate is
strictly zero so that once attached, monomers cannot detach from
clusters. Irreversibility leads to a loss of ergodicity since only a
fraction of the possible cluster configurations will be sampled during
the dynamics, while many others will never be visited.  As a result,
the final ``quenched'' or ``metastable'' cluster size distribution is
not an equilibrium one and depends strongly on initial conditions.
Ergodicity is restored in the case of a non-zero detachment rate,
where all possible cluster configurations are eventually sampled and
where the cluster sizes approach an equilibrium distribution,
independent of the initial configuration. Reversible binding, in the
limit of small unbinding rates, is in analyzed Section IV.
 
Several studies of homogeneous nucleation have shown the existence of
long lived metastable states followed by final equilibration, or
``coarsening'', to a very different cluster size distribution.  These
results were found for nucleating systems driven by an infinite supply
of monomers and while allowing clusters to grow without bound
\cite{WATTIS,PENROSE,SMEREKA}.  Our results for mass-conserving
heterogeneous nucleation show a similar coarsening behavior. The
steady-state cluster distributions arising from the irreversible and
reversible dynamics are very different from each other, especially in
the limit of small particle numbers and even in the case of
vanishingly small detachment rates.  In the latter case, when
unbinding is very slow compared to binding, relaxation to the true
equilibrium cluster size distribution occurs over the long time scales
associated with unbinding. As a result, cluster concentrations reach
long-lived metastable plateaus that depend on initial conditions and
that can be closely approximated by results obtained from considering
irreversible dynamics, as treated in Section III.  Only at longer
times, after monomers start unbinding in appreciable numbers, does
this metastable size distribution ``coarsen'' and cross over to the
true equilibrium one.  While the metastable and equilibrium cluster
distributions are generally very different, we find a surprising
result: for certain sets of parameters, special cluster sizes have
identical metastable and equilibrium concentrations.  For these
clusters, the equilibration process appears to be dramatically
accelerated.  In Section V we find the exact mathematical
relationship leading to the apparent fast coarsening where certain
clusters reach equilibrium concentrations well before the rest. Our
results are a consequence of total mass conservation, and do not arise
in the case of receptors binding an unlimited supply of free ligand
monomers. Finally, in the Conclusions, we discuss implications and
future extensions of our work.

\section{Mass-action equations}

To begin our analysis, we consider a model of heterogeneous nucleation
for $M$ well-mixed monomeric ligands binding sequentially
\cite{PRLDORSOGNA} to any of the $N_{\rm s}$ uniformly dispersed
ligand seeds, neglecting fragmentation and aggregation that do not
involve monomers, since they have been treated in other contexts
\cite{MAJUMDAR,REDNER}.  We also assume that each seed can accommodate
at most $N$ monomers due to stoichiometry constraints and consider the
mean-field mass-action equations for the number of clusters $c_{k}(t)$
of size $k$, where $0 \leq k \leq N$. Here, $k=0$ indicates ``naked
seeds", with no bound monomers, and $k=N$ saturated ones, where no
further binding is possible.  In general, monomer attachment and
detachment rates from a cluster of size $k$ can be explicitly
$k$-dependent. Specific forms for $p_{k}$ and $q_{k}$ have been used
to describe cooperativity and the nucleation of clusters of various
shapes and in different dimensions.  For instance, $p_{k}\sim k^{1/2}$
and constant $q_{k}$ are typically used to model 2D nucleation of
circular droplets when monomer binding is not diffusion-limited
\cite{NIETHAMMER}.  In our work, we assume that while detachment is
independent of the number of particles in the free monomer pool, the
attachment process depends only on how many monomers remain unbound.
The ``Becker-D\"{o}ring'' equations for $c_k(t)$ can thus be written
as

\begin{equation}
\begin{array}{rl}
\dot{c}_{0} & = -p_{0} m(t) c_{0} + q_{1}c_{1},  \\[13pt] 
\dot{c}_{k} &
= -p_{k}m(t)c_{k} - q_{k}c_{k} + p_{k-1}m(t)c_{k-1} + q_{k+1}c_{k+1}, \\[13pt] 
\dot{c}_{N} & = - q_{N}c_{N} + p_{N-1}m(t) c_{N-1}. \\
\end{array} 
\label{ODEHETERO2}
\end{equation}

\noindent
Here, $p_k m(t)$ and $q_k$ represent monomer attachment and detachment
rates, respectively. The attachment rate is proportional to $m(t)$,
the number of free monomers available for binding

\begin{equation}
m(t) \equiv M - \sum_{k=1}^{N} k c_{k}(t).
\label{CONSTRAINT}
\end{equation}

\noindent
Note, that although these equations are written assuming finite particle
numbers, they can also describe concentrations, given a normalizing
reference concentration. The quantities $M$, $N$, $N_{\rm s}$ and
$c_{k}(t)$ therefore need not be integers. We assume a typical initial
condition where all the mass is in the form of monomers, $m(t=0) = M,
c_{0}(t =0) = N_{\rm s}, c_{k>0}(t =0)=0$.  The other constraint
particles must obey is that the total number of seeds must be $N_{\rm
  s}$ at all times, regardless of cluster population levels. We thus
impose

\begin{eqnarray}
\label{constr}
N_{\rm s} = \sum_{j=0}^{N} c_j (t),
\end{eqnarray}

\noindent
which is satisfied by the system in Eq.~\ref{ODEHETERO2}, using the
given initial conditions.  For clarity, and because we will be
referring to these equations often, we rewrite our mass-action
equations for the simplified case of uniform attachment and detachment
rates $p_k = p$ and $q_k = q$. This approximation might be most
relevant for modeling nucleation and growth of linear filaments where
there are always only one or two ends on which monomers can bind or
detach from. Rescaling time by the attachment rate $p$, we find


\begin{equation}
\begin{array}{rl}
\dot{c}_{0} & = -m(t) c_{0} + \varepsilon c_{1}, \\[13pt]
\dot{c}_{k} & =  -m(t)c_{k} - \varepsilon c_{k} + m(t)c_{k-1} + 
\varepsilon c_{k+1}, \\[13pt]
\dot{c}_{N} & = - \varepsilon c_{N} + m(t) c_{N-1},
\end{array}
\label{HETEROEQN}
\end{equation}

\noindent where $\varepsilon = q/p$. In the context of
Eqs.~\ref{HETEROEQN}, irreversible binding corresponds to $\varepsilon
= 0$ and reversible binding to $\varepsilon > 0$. In the following, we
shall be interested in the difference in behavior between $\varepsilon
= 0$ and $\varepsilon \to 0^{+}$. As we shall see, the presence of a
vanishingly small detachment rate $\varepsilon \to 0^+$ can lead to
qualitatively different cluster size distributions compared to those
obtained in the purely irreversible case at $\varepsilon =0$.

\section{Irreversible binding}

We first consider the strictly irreversible binding limit in the
general framework of Eqs.\,\ref{ODEHETERO2} where there is no
detachment and $q_k =0$.  Two possibilities arise. For $M \geq N
N_{\rm s}$, there is an excess of available monomers.  Given the
irreversible nature of the dynamics, all $N_{\rm s}$ seeds will be
fully occupied by $N$ ligands, leaving $M - N N_s$ free ones.  In
this case, we expect steady state solutions to yield $c_N(t \to
\infty) = N_{\rm s}$, $c_{k \neq N}(t \to \infty) = 0$ and $m(t \to
\infty) = M - N N_{\rm s}$. We shall call this the excess monomer
limit.  In the other case of $M < N N_{\rm s}$ there are not enough
monomers to fill seeds to capacity and a non-trivial steady state will
arise. Here, we expect the existence of a finite time $t_*$ at which
the pool of free monomers is depleted, so that $m(t_*) = 0$. At this
time, the final cluster distribution is the one frozen at $t_*$, since
no further attachments nor any detachments are possible. We shall call
this the excess seed limit.  Since the quantity $M-N N_{\rm s}$ will
play an important role in our analysis, we introduce the monomer
excess parameter $\sigma = M / N N_{\rm s}$, so that values of $\sigma
\geq 1$ correspond to the excess monomer case, while $\sigma < 1$
describes the case of excess seeds.

To determine the final cluster size distributions in both cases, first
note that Eqs.~\ref{ODEHETERO2} (or Eqs.~\ref{HETEROEQN}) are
nonlinear due to the constraint on $m(t)$ involving $c_{k}(t)$ via
Eq.~\ref{CONSTRAINT}. If $q_k=0$ however, all terms on the right hand
side multiply $m(t)$.  We can make analytic progress by dividing by
$m(t)$ and defining a rescaled time $\tau$ according to

\begin{eqnarray}
\label{TAU}
\frac{\dd \tau}{\dd t} = m(t) = M - \sum_{k=1}^{N}k c_k(t).
\end{eqnarray}

\noindent Eqs.\,\ref{ODEHETERO2} can now be written as

\begin{equation}
\begin{array}{rl}
\displaystyle{\frac{\dd c_{0}}{\dd \tau}} & = -p_0 c_{0}, \\[13pt]
\displaystyle{\frac{\dd c_{k}} {\dd \tau}} &= p_{k-1} c_{k-1}-p_k c_{k},  \\[13pt]
\displaystyle{\frac{\dd c_{N}}{\dd \tau}} &= p_{N-1} c_{N-1}. \\[13pt]
\end{array}
\label{HETERO0}
\end{equation}

\noindent 
Our goal is to find the rescaled time $\tau_*$ corresponding to 
the rescaled time at which monomers are irreversibly depleted: $m(\tau_*) = M -
\sum_{k=1}^{N} k c_k (\tau_*) =0$.  The quenched, steady-state cluster
size distribution is thus found by evaluating the cluster 
concentrations at $\tau_{*}$: $c_k(\tau_*)\equiv
c^{*}_{k}$. Eqs.\,\ref{HETERO0} are linear and can be solved by using
Laplace transforms.  Upon defining $\tilde c_k (s) = \int_0^{\infty}
e^{-s \tau} c_k(\tau) d\tau$ we find that $\tilde c_k(s)$ satisfy

\begin{equation}
\begin{array}{l}
s \tilde c_0 - N_{\rm s} = - p_0 \tilde c_0, \\[13pt] 
s \tilde c_{k} = - p_{k} \tilde c_k + p_{k-1} \tilde c_{k-1}, 
\\[13pt]
s \tilde {c_N} =  p_{N-1}  \tilde c_{N-1},
\end{array}
\label{ODEHETERO4}
\end{equation}

\noindent which yield the solutions

\begin{eqnarray}
\frac{\tilde c_k(s)}{N_{\rm s}}= \frac{\prod_{j=0}^{k-1} p_j}{\prod_{j=0}^{k}(s+p_j)},
\qquad 
\frac{\tilde c_N(s)}{N_{\rm s}}= \frac{\prod_{j=0}^{N-1} p_j}{s \prod_{j=0}^{N-1}(s+p_j)}.
\label{SOLUTIONPN}
\end{eqnarray}

\noindent
To simplify our analysis, we restrict ourselves to uniform intrinsic
attachment rates $p_k= p$ and use units of time such that $p=1$.  The
dynamics are now described by Eq.\,\ref{HETEROEQN} with $\varepsilon =
0$.  The solutions represented by Eqs.~\ref{SOLUTIONPN} thus simplify to

\begin{eqnarray}
\frac{\tilde c_k (s)}{N_{\rm s}}= \frac 1 {(s+1)^{k+1}}, \qquad 
\frac {\tilde c_N (s)} {N_{\rm s}}= \frac 1 {s (s+1)^N},
\end{eqnarray}

\noindent which can be 
inverse Laplace transformed to yield 

\begin{eqnarray}
\label{results}
\frac {c_{k<N}(\tau)}{ N_{\rm s}} = \frac{ \tau ^k e^{-\tau}}{k!} ,\qquad
\frac {c_N (\tau)} 
{N_{\rm s} } =  1 - \sum_{j=0}^{N-1} \frac {\tau^j e^{-\tau} } {j!}.
\end{eqnarray}

\noindent These results obey the constraint in Eq.\,\ref{constr}.
The value of $\tau_*$ can now be found by using Eqs.~\ref{results} in
the mass constraint Eq.~\ref{CONSTRAINT} and imposing the
condition $m(\tau_*) = 0$.  After some algebra, this condition yields

\begin{equation}
{\tau_{*}^N e^{-\tau_{*}}\over N\Gamma(N)}
+(N-\tau_{*}){\Gamma(N,\tau_{*})\over N\Gamma(N)} = 1 -\sigma.
\label{TRANS}
\end{equation}

\noindent As mentioned above, we expect a finite solution $\tau_*$
only in the excess seed $(\sigma < 1)$ case. For excess monomers
($\sigma \geq 1$) we do not expect a finite time at which monomers are
depleted.  Indeed, the left hand side of the above expression is
positive and monotonically decreasing, implying that Eq.\,\ref{TRANS}
will have a finite, real solution only in the excess seed case, for
$\sigma < 1$. When the initial monomer number $M$ is increased, and
$\sigma$ decreases past unity, the root $\tau_{*}$ diverges since
all binding sites on the seeds are eventually occupied and further depletion
of monomers can never occur. The quenched concentrations in this case
are described by $m(\tau_{*} \to \infty)= M - N N_{\rm s}$ and
$c_k(\tau_{*} \to \infty) = N_{\rm s} \delta_{k,N}$, indicating that
all seeds are filled to capacity for $\sigma \geq 1$, as expected.

As a nontrivial example of the seed excess case, $\sigma < 1$, we can
numerically solve Eq.\,\ref{TRANS} for $N=10$, $M=30$, $N_{\rm s} = 8$
and $\sigma = 3/8$ to obtain $\tau_* = 3.75248$ and 

\begin{equation}
\label{firstsol}
\begin{array}{ll}
\displaystyle \frac {c_0}{N_{\rm s}} = e^{-\tau_*} = 0.02346, & \displaystyle
\frac{c_1}{N_{\rm s}} = \tau_* e^{-\tau_*} = 0.088031, \\
\displaystyle \frac{c_2}{N_{\rm s}} = \frac{\tau_*^{2}}{2!} e^{-\tau_*} = 
0.165168, & \displaystyle
\frac{c_3}{N_{\rm s}} = \frac{\tau_*^{3}}{3!}e^{-\tau_*} = 0.206596, \\
\displaystyle\frac{c_4}{N_{\rm s}} = \frac{\tau_*^{4}}{4!} e^{-\tau_*} = 0.193812, &
\displaystyle \frac{c_5}{N_{\rm s}} = \frac{\tau_*^{5}} {5!} e^{-\tau_*} = 
0.145455, \\
\displaystyle\frac{c_6}{N_{\rm s}} = \frac{\tau_*^{6}}{6!} e^{-\tau_*} = 
0.0909699, & \displaystyle
\frac{c_7}{N_{\rm s}} = \frac{\tau_*^{7}}{7!} e^{-\tau_*} = 0.0487661, \\
\displaystyle \frac{c_8}{N_{\rm s}} = \frac{\tau_*^{8}}{8!} e^{-\tau_*} = 
0.0228743, & \displaystyle
\frac{c_9}{N_{\rm s}} = \frac{\tau_*^{9}}{9!} e^{-\tau_*} = 0.0095373, \\
\displaystyle \frac{c_{10}}{N_{\rm s}} = \frac{\tau_*^{10}}{10!} e^{-\tau_*} = 
0.0035788. & \: 
\end{array}
\end{equation}

\noindent It can be explicitly verified that 
these solutions obey $\sum_{k=0}^{10}c^{*}_{k} = N_{\rm s}$.
Finally, in the case of large $N$, asymptotic analysis of Eq.~\ref{TRANS}
gives

\begin{eqnarray}
\tau_{*} \simeq \frac{M}{N_{\rm s}} + 
\frac{e^{- \frac {M}{N_{\rm s}}} N_{\rm s} }{M \sqrt{2 \pi N}}
\left( \frac{e M}{N N_{\rm s}} \right)^{N}, 
\end{eqnarray}

\begin{figure}[t]
\includegraphics[width=3.4in]{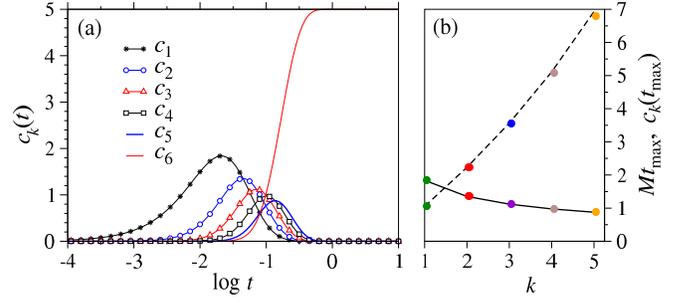}
\caption{(a) Numerical solution to Eqs.~\ref{HETEROEQN} with $M=50$,
  $N=6$, $N_{\rm s}=5$, and $\varepsilon=0$.  Since monomers are in
  excess ($\sigma=5/3>1$), all clusters except $c_{N}$ vanish at long
  times. This plot is indistinguishable from the one plotted using
  $\varepsilon =0.0001$, and is qualitatively similar to what would be
  found for a constant free monomer concentration $m(t) = M$.  (b) The
  numerically computed (colored dots) maximal cluster concentrations
  $c_{k}(t_{\rm max}) = k^{k}e^{-k}/k!$ and corresponding times
  $t_{\rm max}$. The approximation $t_{\rm max} \approx N_{\rm
    s}^{-1}\ln\left[M/(M-N_{\rm s}k)\right]$ and the corresponding
  $c_{k}(t_{\rm max})$ are also shown by the dashed and solid curves,
  respectively.}
\label{FIG2}
\end{figure}

\noindent which allows us to find analytic approximations to the final
quenched values $c_{k}^{*} = c_k(\tau_*)$ in this limit.  The full
irreversible dynamics are illustrated in Fig.\,\ref{FIG2}. In order to
find approximations for the maximal concentrations of clusters
$c_k(t)$ of size $k < N$, we note that from Eq.\,\ref{results} they
occur at the rescaled time $\tau_{\rm max} = k$. To find the
corresponding real time $t_{\rm max}$ we insert Eqs.~\ref{results} into
the scaling relationship Eq.\,\ref{TAU} so that

\begin{eqnarray}
\label{diffeq}
\frac{\dd \tau}{\dd t} = M - N_{\rm s} \sum_{k=1}^{N-1} k \frac{\tau^k
  e^{-\tau}}{k!} - N\left(1 - \sum_{k=0}^{N-1} \frac{\tau^k
  e^{-\tau}}{ k!} \right).
\end{eqnarray}

\noindent Equation \ref{diffeq} can be numerically integrated to 
find $t(\tau)$. However, for large $N$ the first sum on the
RHS can be approximated by its $N \to \infty$
limit

\begin{equation}
\sum_{k=1}^{N-1}k{\tau^{k}e^{-\tau}\over k!}
\approx \tau + O(1/N).
\end{equation}
In this same limit, the second sum is $O (1/N)$.
Eq.~\ref{diffeq} can therefore be accurately approximated by 

\begin{equation}
{\dd \tau\over \dd t} \approx  M - N_{\rm s}\tau,
\end{equation}
which can be analytically integrated to give

\begin{equation}
t(\tau) \approx N_{\rm s}^{-1}\ln\left[M/(M-N_{\rm s}\tau)\right].
\end{equation}
We plot $t_{\rm max}\equiv t(\tau_{\rm max}) = t(k) \approx N_{\rm
  s}^{-1}\ln\left[M/(M-N_{\rm s}k)\right]$ and the associated
$c_{k}(t_{\rm max})=k^{k}e^{-k}/k!$ in Fig.~\ref{FIG2}(b) as a
function of $k$.  The approximation for $t_{\rm max}$ is shown by the
dashed curve in Fig.~\ref{FIG2}(b) and is extremely accurate,
especially for small $k$ where maxima are reached before appreciable
accumulation of larger clusters invalidate the approximation
$\dot{\tau} \approx M-N_{\rm s}\tau$.

In the next section we will analyze the nucleation process when
successive monomer detachment is allowed.  The question will arise as
to how closely the irreversible nucleation results found here are
followed in the case of a vanishingly small, but non-zero, detachment
rate $\varepsilon$.  As we shall see, our reversible results will
closely mirror the irreversible ones in the limit $\varepsilon \to
0^+$, \textit {only} in the excess monomer case, when seeds are
saturated with ligands. In the excess seed case on the other hand,
dramatic differences between reversible and irreversible binding
arise, even as $\varepsilon \to 0^+$. Only very special parameter choices
will lead to the rare matching of reversible and irreversible 
dynamics for specific clusters.

\section{Reversible binding}

In this section we find the equilibrium cluster size distributions
allowing for positive detachment rates $q_k>0$. We start by finding
the equilibrium cluster concentrations $c_k^{\rm eq} \equiv c_k(t \to
\infty)$ by setting $\dd c_{k}/\dd t = 0$ in Eqs.~\ref{ODEHETERO2}.
Due to reversibility, initial conditions are irrelevant. After
defining $m^{\rm eq} \equiv M-\sum_{k=1}^{N}kc_{k}^{\rm eq}$, we find
that $c_{k}^{\rm eq}$ can be written as a function of $c_0^{\rm eq}$
and $m^{\rm eq}$ as follows

\begin{eqnarray}
\label{ck}
c_{k}^{\rm eq} = c_0^{\rm eq} \frac{\prod_{j=0}^{k-1} p_j}
{\prod_{j=1}^{k} q_j} [m^{\rm eq}]^k.
\end{eqnarray}

\noindent This expression can be used in the mass constraint of
Eq.\,\ref{CONSTRAINT} and the total cluster number constraint in
Eq.\,\ref{constr} to find two equations for the two unknowns $c_0^{\rm
  eq}$ and $m^{\rm eq}$

\begin{eqnarray}
\label{one}
m^{\rm eq} &=& M - c_0^{\rm eq} \sum_{k=1}^N k
\frac{\prod_{j=0}^{k-1} p_j} {\prod_{j=1}^{k} q_j}
[m^{\rm eq}]^k, \\
\label{two}
N_{\rm s} &=& c_0^{\rm eq}  \sum_{k=1}^{N} 
\frac{\prod_{j=0}^{k-1} p_j} {\prod_{j=1}^{k} q_j}
[m^{\rm eq}]^k,
\end{eqnarray}

\noindent These equations can be solved by substituting the expression
for $c_0^{\rm eq}$ in Eq.\,\ref{two} into Eq.\,\ref{one} and
determining $m^{\rm eq}$ numerically. Again, computations  are greatly
simplified by restricting our analysis to uniform attachment and
detachment rates $p_k = p$ and $q_k = q$, respectively.  Further
nondimensionalizing time in units of $p^{-1}$ and introducing
$\varepsilon \equiv q/p$, Eqs.~\ref{ck} become

\begin{eqnarray}
c_{k}^{\rm  eq} =c_{0}^{\rm eq} \left[\frac{m^{\rm eq}}{ \varepsilon} \right]^{k} 
\equiv c_{0}^{\rm eq}z^{k},
\end{eqnarray}

%

\noindent
where $z \equiv m^{\rm eq}/\varepsilon$. The fixed seed number constraint
in Eq.\,\ref{constr} yields $c_{0}^{\rm eq}= N_{\rm s}(z-1)/(z^{N+1}-1)$ so that 
by substituting $c_{0}^{\rm eq}z^{k}$ into Eq.~\ref{CONSTRAINT} we find
an equation for $z$:

\begin{equation}
\begin{array}{l}
\displaystyle \left({\varepsilon z\over N_{\rm s}N} - \sigma\right)(z-1)(z^{N+1}-1)+z^{N+2} \\[13pt]
\hspace{2.5cm} \displaystyle -(1+1/N)z^{N+1} + {z\over N} = 0.
\label{POLYHETERO}
\end{array}
\end{equation}
%
%
Eq.~\ref{POLYHETERO} determines the
numerical value for the normalized cluster fugacity $z$.
%
%
In the small detachment limit $\varepsilon\to 0^{+}$, once more, the
two limits of excess monomers and excess seeds naturally arise. In the
excess monomer case, $\sigma \geq 1$, the real root of
Eq.\,\ref{POLYHETERO} can be found as the inner solution of a singular
perturbation \cite{BO} where the largest power of $z$ multiplies
$\varepsilon$ so that $z = \varepsilon^{-1}(M-N_{\rm s}N) + N_{\rm
  s}(M-N_{\rm s}N)^{-1} + O(\varepsilon)$.  Inserting this
approximation for $z$ into the seed constraint Eq.\,\ref{constr} we
find $c_{0}^{\rm eq} \approx N_{\rm s} \left[\varepsilon/(M-N_{\rm
    s}N)\right]^{N} + O(\varepsilon)$, which yields

\begin{equation}
c_{k}^{\rm eq} \approx {N_{\rm s}\over (N_{\rm s}N)^{N-k}} {\varepsilon^{N-k}
\over (\sigma - 1)^{N-k}} + O(\varepsilon^{N-k+1}).
\label{CKEQEM}
\end{equation}

Thus, equilibrium concentrations all vanish as $O(\varepsilon^{N-k})$
except that of the maximum cluster $k=N$ which asymptotes to
$c_{N}^{\rm eq} \approx N_{\rm s} - O(\varepsilon)$.  This qualitative
behavior is expected in the excess monomer case when nearly all
available binding sites are occupied and only nearly fully occupied
seeds survive. In particular, when $\varepsilon\to 0^{+}$, $c_{k \neq
  N}^{\rm {eq}} \to 0$ and $c_{N}^{\rm eq} \to N_{\rm s}$.  This
result is identical to what was found for the strictly irreversible
case of $\varepsilon = 0$ in the previous section.  In the excess
monomer case ($\sigma \geq 1$) thus, all clusters will be filled to
capacity in the case of vanishingly small detachment rates, regardless
of when the limit $\varepsilon \to 0^+$ is taken.

In the opposite case of excess seeds, $\sigma < 1$, monomers are
depleted before all binding sites on all $N_{\rm s}$ seeds can be
filled, leading to finite concentrations $c_{k}^{\rm eq}$.
Interestingly, the excess seed limit {\it further separates into two
  sub-cases}.  From our numerical analysis of Eq.\,\ref{POLYHETERO} we
find that $z >1$ for $1/2 < \sigma <1$, implying $c_{k+1}^{\rm eq} >
c_{k}^{\rm eq}$ and larger cluster sizes tend to be favored.  On the
contrary, for $\sigma < 1/2$ we find $z<1$ so that $c_{k+1}^{\rm eq} <
c_{k}^{\rm eq}$.  In this case, there are too few monomers $M$ for
larger clusters to persist and smaller cluster sizes are more
populated.  For a range of values of $\sigma$ near 1/2 we find that
the approximation


\begin{figure}[t]
\begin{center}
\includegraphics[width=3.2in]{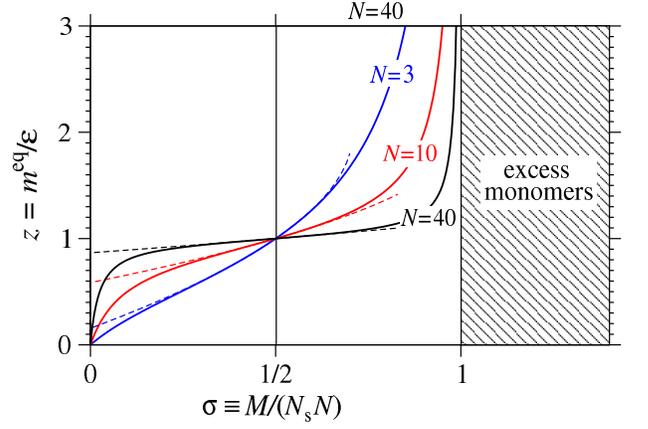}
\caption{Values of the normalized cluster fugacity $z$ determined from
  the real root of Eq.~\ref{POLYHETERO}. Fugacities for $N=3,10,40$
  are plotted in the limit $\varepsilon/(N_{\rm s}N) = 10^{-6}$.  A
  good approximation to $z$ is given by Eq.~\ref{ZAPPROX} and is shown
  by the dashed curves.  The region $\sigma \equiv M/(N_{\rm s}N) \geq 1$
  corresponds to the case of excess monomers where $c_{k}^{\rm eq}\sim
  \varepsilon^{N-k}$, as shown in Eq. \ref{CKEQEM}.}
\label{ZPLOT}
\end{center}
\end{figure}

\begin{equation}
z \approx 2-\left[1-{24\over N+2}\left(\sigma -{1\over 2}\right)\right]^{1/2},
\label{ZAPPROX}
\end{equation}
and the associated $c_{k}^{\rm eq} = c_{0}^{\rm eq}(z)z^{k}$, are
highly accurate.  Note that at the special point $\sigma=1/2$ the
monomer fugacity $z=1$ and all equilibrium concentrations $c_{k}^{\rm
  eq} = N_{\rm s}/(N+1)$ are equal. The behavior of the root $z$ of
Eq.~\ref{POLYHETERO} as a function of the monomer excess is plotted
for $\sigma < 1$ in Fig.~\ref{ZPLOT}. The analytic approximation 
(Eq.~\ref{ZAPPROX}) is also indicated by the dashed curves.





Our analysis thus far does not provide insight into {\it how} the
equilibrium state is reached. As discussed earlier, when $\varepsilon
\to 0^{+}$, we expect binding to occur in a nearly irreversible manner
over intermediate times, yielding metastable cluster size
distributions. Repeated monomer detachment and reattachment become
significant only after much longer times, of the order $t_{\rm c} \sim
\varepsilon ^{-1}$, allowing redistribution of mass into equilibrium
clusters.

To find the metastable cluster size distribution, we make the {\it
  ansatz} that $c_{k}(t)$ can be approximated by setting the
detachment rate $\varepsilon=0$ at intermediate times.  We may thus
neglect detachment and use the results obtained for irreversible
binding up to $t_c \sim \varepsilon^{-1}$, beyond which detachment
effects may become appreciable, both in the excess monomer and excess
seed cases.
%
%

Fig.~\ref{FIG2}(a) shows the full time dependence of $c_{k}(t)$ in the
reversible, excess monomer case where $\sigma = 5/3 \geq 1 $. Here, as
expected, both $c_{k<N}^{\rm eq}$ and $c_{k<N}(\tau\to\infty) \equiv
c_{k<N}^{*}$ vanish as $\varepsilon \to 0 ^+$.  In this case, the
dynamics is not appreciably affected by the onset of detachment and
there are no dramatic behavioral crossovers originating across the
$\varepsilon^{-1}$ time scale.  Reversible and irreversible dynamics
thus coincide at all time scales in the $\varepsilon \to 0^+$ limit if
$\sigma \geq 1$.  In particular, the initial rise in $c_{0<k\leq
  N}(t)$ is determined by the monomer loading process and is
independent of the detachment rate $\varepsilon$.  Setting
$\varepsilon = 0$ and using our results from the previous section we
can numerically compute the time scale $t_{\rm max}$ over which the
cluster size distributions peak.  We have verified that as $\varepsilon
\to 0^+$, the full dynamics arising from the reversible binding
process in not appreciably different from that obtained in the
irreversible binding limit.  In summary, when monomers are in excess,
$\sigma \geq 1$, the differences between $c_{k}^{\rm eq}$ and
$c_{k}^{*}$ vanish in the $\varepsilon \to 0^+$ limit,
ergodicity-breaking is not apparent and seeds are always filled to
capacity.

\begin{figure}[t]
\includegraphics[width=3.4in]{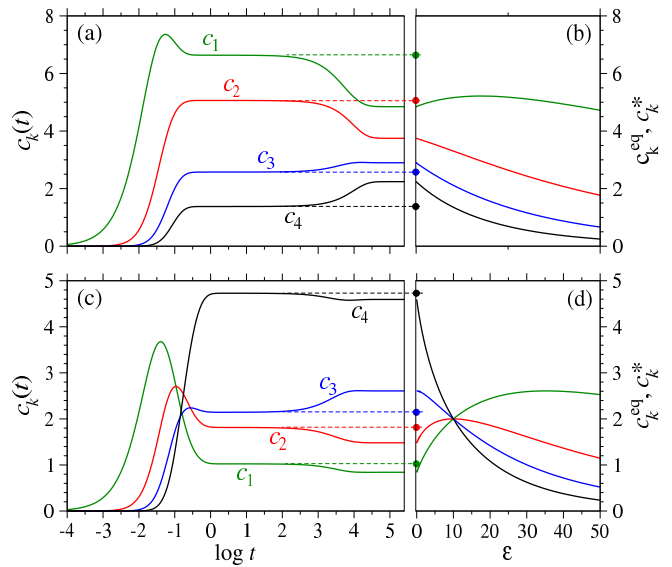}
\caption{Ergodicity breaking occurs only when seeds are in excess.
  (a) Numerical solution to Eqs.~\ref{HETEROEQN} with
  $\varepsilon=0.0001$, $M=30$, $N=4$, and $N_{\rm s}=20$ ($\sigma =
  3/8$).  In this strong excess seed case, $\sigma < 1/2$ and both
  $c_{k}^{*}>c_{k+1}^{*}$ and $c_{k}^{\rm eq} > c_{k+1}^{\rm eq}$.
  (b) $c_{k}^{\rm eq}$ as a function of $\varepsilon$.  Note that even
  as $\varepsilon \rightarrow 0^{+}$, $c_{k}^{\rm eq}$ are different
  from the metastable values $c_{k}^{*}$ (colored dots) found from
  setting $\varepsilon=0$ in Eqs.~\ref{HETEROEQN}.  (c) Cluster
  concentrations $c_{k}(t)$ for $\varepsilon = 0.0001$, $M=30$, $N=4$,
  and $N_{\rm s} = 10$ ($\sigma = 3/4$).  In this weak excess seed
  case, $\sigma > 1/2$, both $c_{k}^{*}<c_{k+1}^{*}$ and $c_{k}^{\rm
    eq} < c_{k+1}^{\rm eq}$. (d) Again, ergodicity-breaking arises
  since $c_{k}^{\rm eq}(\varepsilon \to 0) \not\to
  c_{k}^{*}(\varepsilon\to 0)$.  In all plots, the free monomer
  concentration $m(t)$ and the number of naked seeds $c_{0}(t)$ can be
  reconstructed from the constraint conditions and are not explicitly
  shown.}
\label{FIG3}
\end{figure}

%

We now consider reversible binding in the $\sigma < 1$ case, where
there are more receptor seeds than initial free monomers.  As in the
\textit{ansatz} made in the excess monomer case, we assume that at
least up to time scales of order $\varepsilon ^{-1}$, the dynamics can
be approximated as an irreversible binding process, where $\varepsilon
= 0$.  By following the full dynamics in Eqs.~\ref{HETEROEQN} we
verify that at intermediate times, the metastable concentrations
$c_k(t)$ approach levels approximated by the final ones
$c_{k}(\tau_{*})\equiv c_{k}^{*} \equiv N_{\rm
  s}\tau_{*}^{k}e^{-\tau_{*}}/k!$ reached in the irreversible case
when $\varepsilon = 0$. Following the evolution of
Eqs.\,\ref{HETEROEQN} beyond timescales $\varepsilon^{-1}$, we find
that these metastable concentrations eventually coarsen towards a
qualitatively different, equilibrium distribution defined by
$c_{k}^{\rm eq}(\varepsilon \rightarrow 0^{+})$.

This qualitative difference in cluster size distributions is
noticeable \textit{only} in the excess seed limit when $\sigma < 1$
and illustrates ergodicity-breaking at $\varepsilon = 0$, as clearly
shown in Fig.~\ref{FIG3}.  In Fig.~\ref{FIG3}(a), where $\sigma = 3/8
<1/2$, seeds are in such strong excess that monomers are quickly
depleted and $c_{k+1}^{*}<c_{k}^{*}$. Fig.~\ref{FIG3}(b) plots
$c_{k}^{\rm eq}$ as a function of $\varepsilon$, found from
numerically solving Eq.~\ref{POLYHETERO}. Note that the values of
$c_{k}^{\rm eq}(\varepsilon \to 0^{+})$ differ from the intermediate
ones approximated by the frozen distribution $c_{k}^{*}$.  The latter
are indicated by the colored dots. Figures \ref{FIG3}(c) and (d) are
the analogous plots but for $1/2 < \sigma < 1$, where
$c_{k+1}^{*}>c_{k}^{*}$. When seeds are in excess, the crossover to
equilibrium is clearly observable over the coarsening time scale
$t_{\rm c} \sim \varepsilon^{-1}$.


\section{Apparent accelerated equilibration of specific clusters}

The general qualitative behavior described in the previous section is
that when seeds are in excess ($\sigma < 1$), the full heterogeneous
nucleation problem exhibits dynamics occuring over two time scales.
The first is of $t \sim O(1)$ and corresponds to monomer attachment
rates, while the second coarsening time scale, $t_{\rm c} \sim
\varepsilon^{-1}$, is associated with the monomer detachment rate. In
general, $c^{\rm eq}_{k} \neq c_k^*$.
 
However, upon fine tuning relevant parameters, we find special values
of $\sigma<1$ and $N$ where up to two specific cluster sizes $k$ can
have nearly equal values of quenched and equilibrium concentrations
($c_{k}^{\rm eq} \approx c_{k}^{*}$) in the $\varepsilon \to 0^{+}$
limit.  These clusters quickly reach their equilibrium concentrations
on a short time scale independent of $\varepsilon$. Mathematically,
the sizes $k$ that are subjected to this rapid, apparent equilibration
can be found by setting $c_{k}^{*}=c_{k}^{\rm eq}$:

\begin{equation}
N_{\rm s}{\tau_{*}^{k}e^{-\tau_{*}}\over k!} = N_{\rm s}{(z-1)z^{k}\over (z^{N+1}-1)},
\end{equation}
where $\tau_{*}$ and $z$ are determined by
Eqs.~\ref{TRANS} and \ref{POLYHETERO}, respectively. 

\begin{figure}[t]
\includegraphics[width=3.4in]{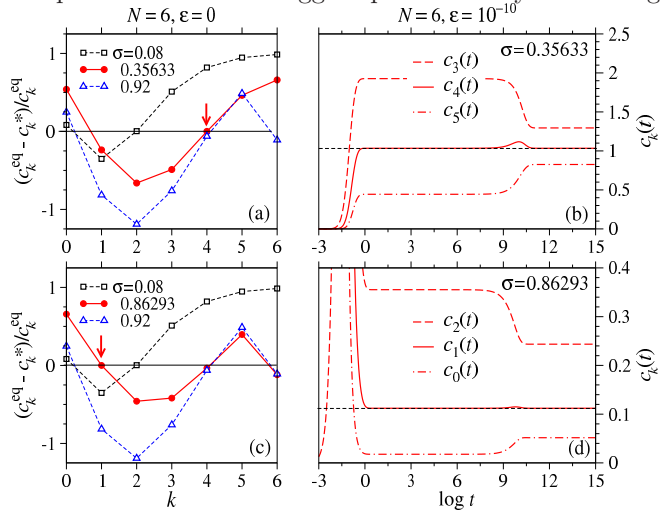}
\caption{(a) The relative difference $(c^{*}_{k} - c_{k}^{\rm
    eq})/c_{k}^{\rm eq}$ as a function of discrete values of $k$ for
  $\sigma=0.08,0.35633, 0.92$. (b) Selected cluster concentrations for
  $\sigma \equiv M/(N_{\rm s}N) = 0.35633$, where $c^{*}_{k=4} \approx
  c_{k=4}^{\rm eq}$. Note the small transient in $c_{4}(t)$ near $t
  \sim \varepsilon^{-1}$. (c) $(c^{*}_{k} - c_{k}^{\rm eq})/c_{k}^{\rm
    eq}$ plotted as a function of $k$ for $\sigma=0.08,0.86293,0.92$
  These values indicate that for $\sigma=0.86293$, the concentration
  $c_{1}(t)$ quickly reaches its equilibrium value. (d) The
  corresponding concentration plot showing just $c_{0}(t), c_{1}(t)$,
  and $c_{2}(t)$. These plots also indicate that for $\sigma = 0.08$,
  the concentration $c_{2}(t)$ experiences accelerated
  equilibration.}
\label{FIG4}
\end{figure}

Figure \ref{FIG4}(a) shows the relative difference
$(c_{k}^{*}-c_{k}^{\rm eq})/c_{k}^{\rm eq}$ for fixed $N=6$ and
$\sigma$ as a function of the discrete cluster size $k$.  Generally,
we find that many values of $0 < \sigma < 1$ give rise to at least one
value of $k$ at which quenched cluster concentrations equal
equilibrium concentrations. In the above example, $N=6$ and $\sigma =
0.35633$, and clusters of size $k=4$ (red arrow) quickly quench to
their equilibrium values. Fig.~\ref{FIG4}(b) plots the numerical
solution to Eq.~\ref{HETEROEQN} for $\varepsilon = 10^{-10}$, $N=6$
and $\sigma = 0.35633$. For simplicity, we have plotted only
$c_{3}(t), c_{4}(t)$, and $c_{5}(t)$.  Note that even though
$c_{4}^{*} = c_{4}^{\rm eq}$ (dashed line), $c_{4}(t)$ does suffer a
small transient perturbation due to the rearrangement of all other
clusters $c_{k\neq 4}$ at time $t \sim \varepsilon^{-1}$ temporarily
disturbing the balance of $c_{4}(t)$.  Figs.~\ref{FIG4}(c) and (d)
illustrate the behavior for $\sigma = 0.86293$. Here,
Fig.~\ref{FIG4}(c) predicts that $c_{1}$ quickly reaches its
equilibrium value. Fig.~\ref{FIG4}(d) explicitly plots $c_{0}(t),
c_{1}(t)$, and $c_{2}(t)$ for $\sigma = 0.86293$.  Figs.~\ref{FIG4}(a)
and (c), also suggest that $c_{4}^{*} \approx c_{4}^{\rm eq}$ over a
wide range of values of $\sigma$.  For the finite processes we have
considered, we find that at most two sizes $k$, out of $N$, can
exhibit accelerated equilibration, provided $\sigma$ and $N$ are
precisely tuned. We expect the qualitative aspects of these results to
hold when binding and/or unbinding rates are weakly cluster
size-dependent, allowing our analysis to apply in scenarios of weakly
cooperative ligand-receptor binding \cite{PRLDORSOGNA}.

%

\section{Conclusions}

In this work we have analyzed a simple and mathematically tractable
model of heterogeneous nucleation to describe ligand-receptor binding
in closed systems, such as within cells or organelles.  A complete
analysis in terms of the parameters $(M,N,N_{\rm s})$ in the
$\varepsilon \to 0^+$ limit shows how dramatically differently the
system behaves in a monomer-rich environment compared to a seed-rich
one.  In the latter case, when binding sites outnumber initial
monomers ($\sigma < 1$), we find that after an initial transient $\sim
t_{\rm max}$, cluster densities first approach $c_{k}^{*}$,
approximating the quenched concentrations when $\varepsilon=0$ and all
free monomers have been depleted. This long-lived metastable
distribution eventually coarsens to a very different equilibrium
distribution $c_{k}^{\rm eq}$ at much later times $t_{\rm c} \sim
1/\varepsilon$. Surprisingly, when parameters ($\sigma$ and $N$) are
finely tuned, it is also possible that $c_{k}^{*}\approx c_{k}^{\rm
  eq}$, for particular $k-$clusters, resulting in much shorter
coarsening times.  We find that clusters of up to two specific sizes
may appear to reach their equilibrium concentrations by $t \sim
(1)$. Our results have general implications for ligand-receptor
kinetics and suggest practical ways of tuning $(M,N,N_{\rm s})$ in
experiments to accelerate the equilibration of specific clusters by
stabilizing their metastable sizes.

A number of extensions of our analysis can be further investigated.
Certain forms for cluster size-dependent attachment and detachment
rates, $p_{k}$ and $q_{k}$, can be incorporated into the analysis.
For example, if $p_{k}\sim k$, certain products and sums in
Eqs.~\ref{ck}, \ref{one}, and \ref{two} can be analytically expressed
or approximated to derive variations to Eq.~\ref{POLYHETERO} and the
associated concentrations $c_{k}^{\rm eq}$. Furthermore, for small
numbers of clusters, the mean-field results derived from the
Becker-D\"{o}ring equations may deviate from the expected cluster size
distributions arising from fully stochastic simulations
\cite{BHATT2003}. We expect our mean-field results to be qualitatively
valid when cluster correlations are included in the dynamics. A
careful quantitative investigation of stochastic effects will be
included in future work.

\vspace{3mm}

This work was supported by the National Science Foundation through
grants DMS-0719462 (MD), DMS-1021850 (MD), DMS-1032131 (TC), and
DMS-1021818 (TC), and Army Research Office through grant 58386MA (TC).

\end{document}